# Least-squares fitting of Gaussian spots on graphics processing units

Marcel Leutenegger[1*], Michael Weber[1]

[1] Max Planck Institute for Biophysical Chemistry, Department of NanoBiophotonics, Am Faßberg 11, 37077 Göttingen, Germany.

[*] Corresponding author
E-mail: marcel.leutenegger@alumni.epfl.ch

The investigation of samples with a spatial resolution in the nanometer range relies on the precise and stable positioning of the sample. Due to inherent mechanical instabilities of typical sample stages in optical microscopes, it is usually required to control and/or monitor the sample position during the acquisition. The tracking of sparsely distributed fiducial markers at high speed allows stabilizing the sample position at millisecond time scales. For this purpose, we present a scalable fitting algorithm with significantly improved performance for two-dimensional Gaussian fits as compared to Gpufit.

## Introduction

On the single-molecule level, optical nanoscopy can now achieve nanometer-scale spatial resolution upon imaging the fluorophore distribution in biological or synthetic samples [1–5]. However, due to the inherently extended image acquisition times, typically many minutes, it is mandatory to keep track of the sample position with even better accuracy for an active stabilization of the sample position during the acquisition and/or drift correction of the measured fluorophore positions.

To avoid interference with the excitation and detection spectra of the imaged fluorophores and to limit the exposure of the sample to potentially harmful light, the focus and sample tracking often uses the near-infrared spectrum from the 850 to 1050 nm wavelength range. In this range, CMOS cameras are still sensitive enough and readily available. Variations in the pixels' control and read-out circuits can be greatly reduced by careful calibration [6]. However, in the near-infrared range optical etaloning and variations in the temperature, the thickness and/or the surface quality of many sensors still affect the detection efficiency and produce notable tracking errors. An active stabilization of the sample position greatly reduces the issue with distinct pixel responses and is thus key for sub-nanometer accuracy.

Fast CMOS or scientific CMOS cameras acquire images at hundreds to thousands of frames per second (fps), whereas each image may contain the spots of dozens of fiducial markers. The simultaneous tracking of these markers and the averaging of their positions among several images is usually required to achieve the desired accuracy. Hence, the active stabilization of the sample position demands for an accurate and reliable localization of a sufficient number of fiducial markers in real time. For this purpose, we report on a massively scalable algorithm for the least-squares fitting of image spots with approximately two-dimensional Gaussian profiles. The algorithm was inspired by Gpufit [7] but specializes in fits of image spots with at most 1024 pixels in order to boost the performance of the calculations. This specialization allows keeping most intermediate results in the registers of NVIDIA graphics processing units (GPUs), such that slow memory access is minimized. Moreover, after data transfer to the GPU, the fitting algorithm runs on the GPU without further intervention. Compared to Gpufit, these choices improved the throughput by at least 35 % and often by 100 % or more.

## Method

Fitting of image spots with a two-dimensional profile and uniform background requires at least five parameters: two for the coordinates of the spot location; one for the spot size; one for the spot intensity; and one for the background intensity. However, the number of independent parameters can be reduced by determining the intensities implicitly [8], as shown below.

### Implicit-amplitudes least-squares fitting

Given the pixel positions $\vec{r}_i$, the parameters $\vec{p} = [p_1, \ldots, p_j]^T$ of an image model function $h(\vec{p})$: $h_i(\vec{p}) = h(\vec{r}_i, \vec{p})$ to fit an image spot $g$: $g_i = g(\vec{r}_i)$ can be estimated by minimizing the squared error $\chi^2(\vec{p})$:

$$\underset{\vec{p}}{\mathrm{argmin}}\left(\chi^2(\vec{p}) = \sum_{i=1}^N (h_i(\vec{p}) - g_i)^2\right) \quad (1)$$

In general, the model function includes amplitudes for the background and the shape-defining profile to fit the image spot. In the presence of locally varying background, the background structure may be estimated and subtracted beforehand [9]. In the following, we assume that the captured image contains a uniform background $\beta$ plus the signal $\alpha f(\vec{p})$ obtained by scaling the normalized profile $f(\vec{p})$: $f_i(\vec{p}) = f(\vec{r}_i, \vec{p})$ by its amplitude $\alpha$. Therefore, we defined the image model function as $h(\vec{p})$: $h_i(\alpha, \beta, \vec{p}) = \alpha(\vec{p})f(\vec{r}_i, \vec{p}) + \beta(\vec{p})$ and minimized the squared error for the reduced set of shaping parameters $\vec{p}$:

$$\underset{\vec{p}}{\mathrm{argmin}}\left(\chi^2(\vec{p}) = \sum_{i=1}^N (h_i(\alpha, \beta, \vec{p}) - g_i)^2\right) \quad (2)$$

The calculation of the modified model function now requires three steps:

1. Evaluate the profile $f_i(\vec{p})$ for the shaping parameters $\vec{p}$.

2. Estimate the coefficients $\hat{\alpha}(\vec{p})$ and $\hat{\beta}(\vec{p})$ by implicitly minimizing $\chi^2(\alpha, \beta \mid \vec{p})$:

$$\underset{\alpha,\beta}{\mathrm{argmin}}\left(\chi^2(\alpha, \beta \mid \vec{p}) = \sum_{i=1}^N (\alpha f_i(\vec{p}) + \beta - g_i)^2\right) \quad (3)$$

3. Evaluate the model function $h_i(\hat{\alpha}, \hat{\beta}, \vec{p}) = \hat{\alpha}(\vec{p})f_i(\vec{p}) + \hat{\beta}(\vec{p})$.

In step 2, the estimates of the coefficients $\hat{\alpha}$ and $\hat{\beta}$ were obtained by setting the error gradient to zero:

$$\nabla \chi^2(\alpha, \beta \mid \vec{p}) = \begin{bmatrix} \frac{\partial \chi^2}{\partial \alpha} \\ \frac{\partial \chi^2}{\partial \beta} \end{bmatrix} = 2 \sum_{i=1}^N \begin{bmatrix} (\alpha f_i + \beta - g_i)f_i \\ \alpha f_i + \beta - g_i \end{bmatrix} = 0 \quad (4)$$

For brevity, we define

$$F := \sum_{i=1}^N f_i \qquad G := \sum_{i=1}^N g_i \qquad \mathbf{F} := \sum_{i=1}^N f_i^2 \qquad \mathbf{G} := \sum_{i=1}^N f_i g_i \quad (5)$$

and solved for the coefficients by zeroing the partial derivatives of $\chi^2$:

$$\begin{bmatrix} \hat{\alpha}\mathbf{F} + \hat{\beta}F = \mathbf{G} \\ \hat{\alpha}F + \hat{\beta}N = G \end{bmatrix} \quad \rightarrow \quad \hat{\alpha} = \frac{N\mathbf{G} - FG}{N\mathbf{F} - F^2} \quad \hat{\beta} = \frac{G\mathbf{F} - F\mathbf{G}}{N\mathbf{F} - F^2} \quad (6)$$

As there is exactly one solution if $N\mathbf{F} \neq F^2$, which is guaranteed for non-constant strictly positive or negative functions, it must be the minimum.

For finding the shaping parameters that minimize the error, we apply the modified Levenberg–Marquardt algorithm to approach the error minimum iteratively. The algorithm requires the partial derivatives of $\chi^2(\vec{p})$ with respect to the shaping parameters $p_j$. For brevity, we write

$$\partial_j F = \sum_{i=1}^{N} \frac{\partial f_i}{\partial p_j} \qquad \partial_j \mathbf{F} = 2\sum_{i=1}^{N} f_i \frac{\partial f_i}{\partial p_j} \qquad \partial_j \mathbf{G} = \sum_{i=1}^{N} g_i \frac{\partial f_i}{\partial p_j} \qquad \gamma_j = N\,\partial_j \mathbf{F} - 2F\,\partial_j F \quad (7)$$

$$\frac{\partial \hat{\alpha}}{\partial p_j} = \frac{N\,\partial_j \mathbf{G} - G\,\partial_j F - \hat{\alpha}\gamma_j}{N\mathbf{F} - F^2} \qquad \frac{\partial \hat{\beta}}{\partial p_j} = \frac{G\,\partial_j \mathbf{F} - \mathbf{G}\,\partial_j F - F\,\partial_j \mathbf{G} - \hat{\beta}\gamma_j}{N\mathbf{F} - F^2} \quad (8)$$

and obtain for the partial derivatives of the squared error $\chi^2(\vec{p})$:

$$\frac{\partial \chi_i^2}{\partial p_j} = 2(\hat{\alpha} f_i + \hat{\beta} - g_i)\left(\frac{\partial \hat{\alpha}}{\partial p_j} f_i + \hat{\alpha}\frac{\partial f_i}{\partial p_j} + \frac{\partial \hat{\beta}}{\partial p_j}\right) \quad (9)$$

## Two-dimensional Gaussian model

The shape of a rotationally symmetric two-dimensional Gaussian model is defined by three parameters $\vec{p} = [\bar{x}, \bar{y}, \sigma]^\mathrm{T}$, where $(\bar{x}, \bar{y})$ are the center coordinates and $\sigma$ is the standard deviation of the peak.

$$f_i(\vec{p}) = \exp\left(-\frac{(x_i - \bar{x})^2 + (y_i - \bar{y})^2}{2\sigma^2}\right) \quad (10)$$

Equation (9) requires its partial derivatives, which are immediately obtained:

$$\frac{\partial f_i}{\partial \bar{x}} = \frac{x_i - \bar{x}}{\sigma^2} f_i \qquad \frac{\partial f_i}{\partial \bar{y}} = \frac{y_i - \bar{y}}{\sigma^2} f_i \qquad \frac{\partial f_i}{\partial \sigma} = \frac{(x_i - \bar{x})^2 + (y_i - \bar{y})^2}{\sigma^3} f_i \quad (11)$$

Similar to Gpufit, Fit2DGaussian allocates one GPU thread per pixel for all calculations. In contrast to additions and multiplications, divisions and exponentials are computationally expensive operations that run only on few GPU threads in parallel. Access to shared memory is equally slow, though. Therefore, each thread inverts $\sigma$ once and takes the exponential once per evaluation of the profile function and its partial derivatives.

## Levenberg-Marquardt algorithm

In the following, we collate the values at the different sampling positions in column vectors, for instance $\vec{h} = [h_1, \ldots, h_i, \ldots, h_N]^\mathrm{T}$. To minimize the squared error, we used the scale-invariant damped Levenberg-Marquardt algorithm [10,11,12]. The parameter estimates are updated iteratively ($\vec{p}' = \vec{p} + \vec{\delta}$) until convergence ($\vec{\delta} \to 0$ or $\chi^2 \to 0$) or no further reduction of the error ($\chi'^2 > \chi^2$) is obtained. At each iteration, the parameter update step $\vec{\delta}$ is obtained by solving

$$\left(\mathbf{J}^\mathrm{T}\mathbf{J} + \lambda\,\mathrm{diag}(\mathbf{J}^\mathrm{T}\mathbf{J})\right)\vec{\delta} = \mathbf{J}^\mathrm{T}(\vec{h} - \vec{g}), \quad (12)$$

where $\lambda$ is a damping factor and the Jacobian is

$$J = \left[\frac{\partial \overrightarrow{\chi^2}}{\partial p_1}, \ldots, \frac{\partial \overrightarrow{\chi^2}}{\partial p_j}\right]. \tag{13}$$

As there are only three shaping parameters in our model, $J^T J$ is a $3 \times 3$ matrix and equation (12) is trivial to solve analytically and implement efficiently.

$$\vec{\delta} = \left(J^T J + \lambda \operatorname{diag}(J^T J)\right)^{-1} \left(J^T (\vec{h} - \vec{g})\right) \tag{14}$$

The damping factor is initially set to 0.01 and adjusted in each iteration. If the error was reduced, the damping is reduced as well ($\lambda' = \lambda/10$). Otherwise it is increased ($\lambda' = 10\lambda$) until the error decreases or the parameters' change $\vec{\delta}$ is less than the requested minimum. If the damping exceeds $10^4$, the iteration is terminated with an indication that the fit did not converge.

## Implementation

NVIDIA GPUs feature a hierarchical scalable programming model for the massively parallel execution of many threads with similar – at best identical – instruction sequences. The threads are organized in thread blocks and executed by a multiprocessor in groups of 32 threads, called warps. Each multiprocessor can schedule and run several warps simultaneously. The multiprocessors of a GPU board can execute hundreds to thousands of threads simultaneously and provide a massive boost in calculation throughput. For a detailed description of the architecture of NVIDIA GPUs and the CUDA programming model, please refer to the documentation [13].

The threads of a warp can exchange data directly among registers by warp shuffle operations. The threads of a thread block can access shared memory on the multiprocessor to exchange data. Dedicated instructions are available for synchronizing threads in a warp or thread block. Global memory is required for storing large data sets; for exchanging data between thread blocks; and for synchronizing thread blocks on GPUs without dynamic parallelization capability.

The GPU board features a large global memory that is used to exchange data with the host computer. In addition, each multiprocessor features shared memory (typically 64 kilobytes) and a large register file (typically $2^{16}$ 32 bit registers). Whereas access to registers is always immediate, shared memory and in particular global memory accesses are slower and should be organized in particular patterns to best utilize the memory bandwidth.

Accounting for the characteristics of GPU boards and its multiprocessors, the fitting algorithm allocates one thread per pixel; 4 bytes of shared memory and up to 25 registers per thread; an integer number of warps and 92 bytes of shared memory per fit; up to 6 fits per thread block; and enough blocks for all fits. The pixel values, parameters, model values and calculation intermediates are all 32 bit floating point values and stored in memory as such. Figure 1 illustrates this GPU resource mapping for fitting 5 images of $9 \times 9$ pixels with 2 fits per thread block.

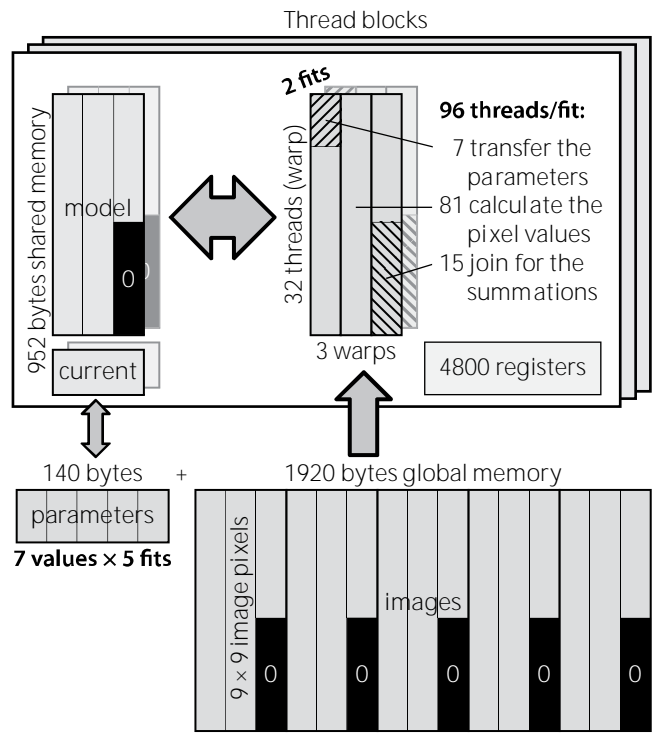

Figure 1. Illustration of the GPU resource mapping for fitting 5 images of 9 × 9 pixels each. The thread blocks run two fits each. The pixel values and parameters are stored in global memory as 32 bit floating point values.

The shared memory per thread is used for sums and dot products of values calculated by the threads. The shared memory per fit caches the current and best parameter sets as well as other intermediate results required by all threads of a fit, for instance the amplitudes $\alpha$ and $\beta$. The host receives these parameters and fit results: stop condition and number of remaining iterations; $\hat{x}$; $\hat{y}$; $\hat{\sigma}$; $\hat{\alpha}$; $\hat{\beta}$; and normalized $\chi^2$.

Running each fit on whole warps allows fully coalesced accesses to global memory and bypasses some shared memory accesses by warp shuffle operations. The number of pixels per fit was limited to the thread block size (1024) to minimize the thread synchronization overhead. Although each fit includes partially idling threads, the lower utilization of the GPU was more than compensated by the simplified calculations.

The core of the **Fit2DGaussian** kernel is summarized in pseudo-code, where the definition operator indicates memory access. Fit2DGaussian gets pointers on the images and parameters data in global device memory (images, parameters); the stop criteria (maxError, minDelta, minStep); the width and the number of pixels of the images (width, pixels); and the number of fits. It initializes registers containing the pixel and fit indices (pixel, fit); the pixel coordinates (x, y); pointers on the individual values in global and shared memory (image, model, current); and it copies the initial parameter values into shared memory. Fit2DGaussian then runs the Levenberg-Marquardt iterations by calling **Gaussian2D** and **SolveForStep** repeatedly until any of the stop criteria is reached. Fit2DGaussian then calculates the normalized squared error and writes the results back to the global memory.

```
Fit2DGaussian<<<blocks,threads,shared>>>(images,parameters,maxError,minDelta,minStep,width,pixels,fits)
{   (pixel,fit,x,y) = from(block,thread,width,pixels);
    if (fit >= fits) Stop;
    image = images(pixel,pixels,fit);                              // initialize memory pointers
    model = shared(thread);
    current := parameters;                                         // read parameters
    lambda = 0.01;
    while(current.status > 0)                                      // iterate until stop condition reached
    {   current.status = current.status – 1;
        chi = Gaussian2D(image,model,current,x,y,pixel,pixels,true); // residuals and derivatives
        if (isnan(chi)) break(current.status := NotConverged);     // stop on error
        if (chi < maxError) break(current.status := MaxError);     // too small a residuum
        best := current;                                           // backup parameters
        step := SolveForStep(current,lambda,pixel);                // get parameter change
        current := limit(best + step);                             // update parameters
        Chi = Gaussian2D(image,model,current,x,y,pixel,pixels,false); // residuals
        if (chi > Chi) lambda = lambda/10;                         // decrease damping
        while(notall(|step| < minStep) and chi < Chi and lambda < 10^4)
        {   current := best;                                       // restore parameters
            lambda = 10*lambda;                                    // increase damping
            step := SolveForStep(current,lambda,pixel);            // get parameter change
            current := limit(best + step);                         // try smaller update
            chi = Gaussian2D(image,model,current,x,y,pixel,pixels,false);
        }
        if (chi < Chi or isnan(Chi))
        {   current := best;                                       // restore parameters
            break(current.status := NotConverged);                 // stop on error
        }
        if (Chi < maxError) break(current.status := MaxError);     // too small a residuum
        if (chi×(1 – minDelta) < Chi) break(current.status := MinDelta); // too small an improvement
        if (all(|step| < minStep)) break(current.status := MinStep); // too small an update
    }
    current.Chi := NormalizedChi(image,model,current,x,y,pixel,pixels); // normalized residuum
    parameters := current;                                         // write parameters
}
```

**Gaussian2D** calculates the model function and its partial derivatives. It is summarized below in pseudo-code. **AlphaBeta** evaluates equation (6). **Sum** cumulates values of all threads per fit and (12) **SolveForStep** evaluates equation (14). Please refer to the source code and its enclosed documentation for details.

```
Chi = Gaussian2D(image,model,current,x,y,pixel,pixels,gradient)
{   ex = exp(–0.5*((x – current.x)² + (y – current.y)²)/current.sigma²);
    (current.alpha,current.beta) := AlphaBeta(image,model := ex,current,pixel,pixels);
    nu = image – (current.alpha×ex + current.beta);
    Chi = Sum(model := nu²,pixel,pixels);
    if (gradient)
    {   current.Jacobian :=
        [   Sum(model := nu*∂Chi/∂x,pixel,pixels);
            Sum(model := nu*∂Chi/∂y,pixel,pixels);
            Sum(model := nu*∂Chi/∂sigma,pixel,pixels)
        ];
        current.Hessian := current.Jacobianᵀ · current.Jacobian;
    }
}
```

## Results and Discussion

We measured the performance of our implementation and compared it with the corresponding Gpufit model for 2D Gaussian profiles including a uniform background. For this purpose, we simulated Gaussian image spots in images of $S \times S$ pixels width and height. The spots' centers were placed with a normal distribution of $S/20$ standard deviation from the image centers. The spots' standard deviations were uniformly distributed between 1 and 2. The Gaussian peak amplitude was defined by the integrated signal $N_{\text{signal}} = 400$ counts of the Gaussian profile. The background per pixel was set to $S^{-2}$ of the total background $N_{\text{background}} = 40$ counts. We added Gaussian noise of variance $g_i$ to the calculated image intensities $g_i$ and rounded the results to non-negative integers. The obtained values were approximately Poisson distributed.

We then measured the execution times for image sizes $S \in \{4,5,...,32\}$ in batches of 10 to 10'000 fits per launch of the GPU fitting kernel. The batches of 10, 100, 1'000 and 10'000 fits were run 200×, 20×, 10× and once, respectively. The execution time of each run measured the transfer of the data from the host memory to the GPU, the launch and execution of the kernel, and the retrieval of the fit results back in host memory. It also includes the overhead of the MATLAB statements for repeatedly calling the functions Fit2DGaussian and Gpufit but excludes the estimation of the initial parameters.

For better comparability, the initial parameters were estimated once and passed identically to both functions. The image was smoothed by a moving average with $3 \times 3$ pixels window size. The center position $(x, y)$ was then initialized to the smoothed image pixel with the highest value; the background $\beta$ to the lowest pixel value; and the signal amplitude $\alpha$ to the highest pixel value minus $\beta$. The standard deviation was initialized to $\sigma = \sqrt{M/\pi}$, where $M$ was the number of original image pixels whose value exceeded $\alpha \exp(-0.5) + \beta$.

We limited Fit2DGaussian to 20 iterations and applied the following criteria to stop earlier: (i) residuum $\chi^2 < A/3$ (neither weighted nor normalized); (ii) decrease of $\Delta \chi^2 < 10^{-6} \chi^2$; or (iii) change of all parameters $|\Delta p_j| < 10^{-4} |p_j|$. Gpufit was equally limited to 20 iterations and a parameter tolerance of $10^{-4}$ was applied to stop earlier.

Fit2DGaussian was compiled with the NVIDIA CUDA compiler version 10.1 and benchmarked on NVIDIA GPUs with display driver version 425.25 and compute capability 3.0 and 6.1 using the same assembly code (PTX file). If not specified otherwise, the measurements were performed on an inexpensive NVIDIA Quadro P600 graphics card (compute capability 6.1) and by using MATLAB R2018b running on a Windows 10 computer in the native 64 bit environment.

### Throughput and latency of fits

In this section, the algorithms were benchmarked for 400 signal and 40 background counts.

Figure 2 shows the average execution time per call of the Fit2DGaussian MATLAB function. The execution time is dominated by an overhead of about 0.5 ms for small batches and image sizes. Otherwise, it scales proportionally with the number of fits and pixels per image. Fits of images with $8 \times 8$ pixels and in particular $16 \times 16$ pixels match best with the GPU architecture and run 20 to 50 % faster than for slightly smaller and larger images.

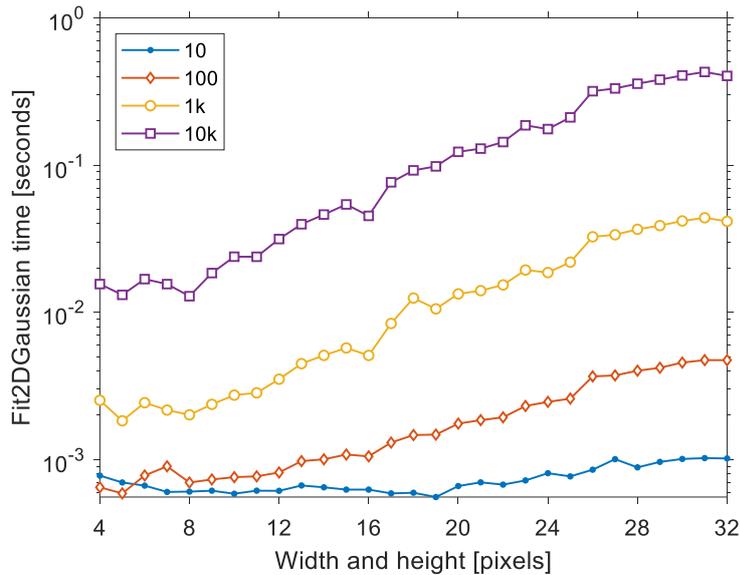

Figure 2. Execution time per call of Fit2DGaussian in seconds for batches of 10 to 10'000 fits per call.

Figure 3 shows the number of fits Fit2DGaussian performs per second. In single-molecule localization microscopy, images usually span 50 to 150 pixels (7 to 12 pixels image width and height). With these dimensions, Fit2DGaussian delivers about 100'000 fits per second with a moderate batch size of 100 fits and about 400'000 fits per second with batch sizes of 1000 fits or more.

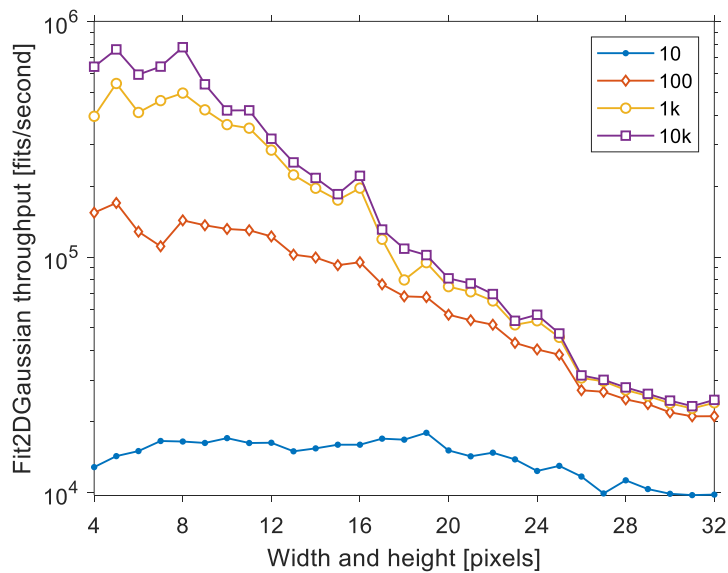

Figure 3. Performance of Fit2DGaussian in fits per second for batches of 10 to 10'000 fits per call.

Figure 4 illustrates the performance of Fit2DGaussian by the number of pixel values that were fitted per second. It shows that our Quadro P600 board levels off at about 40 Million pixels per second. For image extents $S \leq 25$ pixels at least four thread blocks can run concurrently per multiprocessor, otherwise the multiprocessor can only serve two or three thread blocks because of register exhaustion.

 

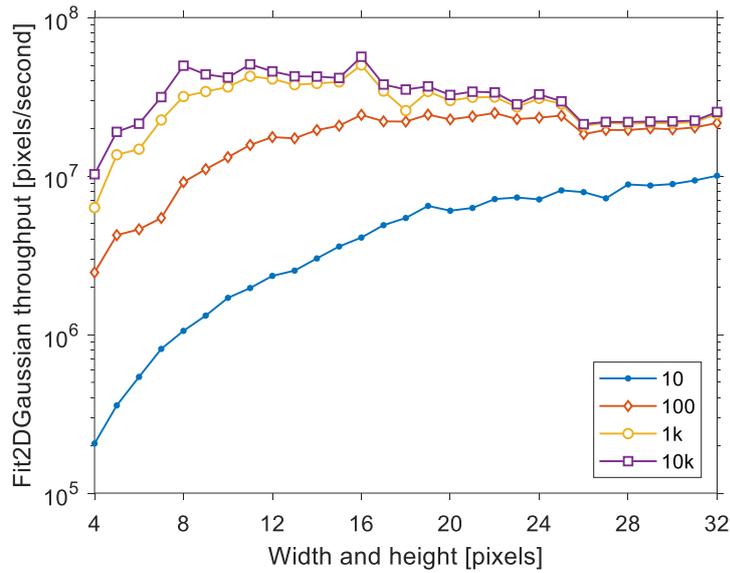

Figure 4. Performance of Fit2DGaussian in fitted pixels per second for batches of 10 to 10'000 fits per call.

The reduced overall calculation performance for $S \geq 26$ pixels might by caused by exposed resource latencies that cannot be hidden by executing other threads. As each fit occupies at least one warp, the performance for $S \leq 7$ pixels is reduced by the noticeable fraction of idling threads.

A comparison of the execution times of Fit2DGaussian with respect to Gpufit is shown in Figure 5. The modified least-squares algorithm and our implementation improve the performance by at least 35 % even in the most favorable conditions for Gpufit, i.e. large batches and large images. With fewer fits and/or smaller images, our implementation doubles, triples or even quadruples the performance.

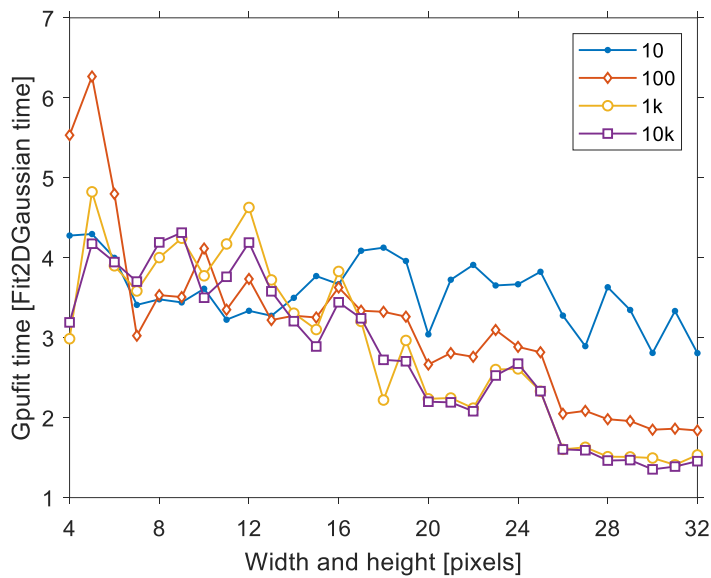

Figure 5. Performance improvement of Fit2DGaussian over Gpufit for batches of 10 to 10'000 fits.

As Figure 6 illustrates, on a NVIDIA Quadro K600 board with compute capability 3.0, Fit2DGaussian at least doubles the performance as compared to Gpufit and outperforms it by an order of magnitude in the most interesting range of image sizes. The slower access to the memory of this card clearly penalizes Gpufit for its greater memory use.

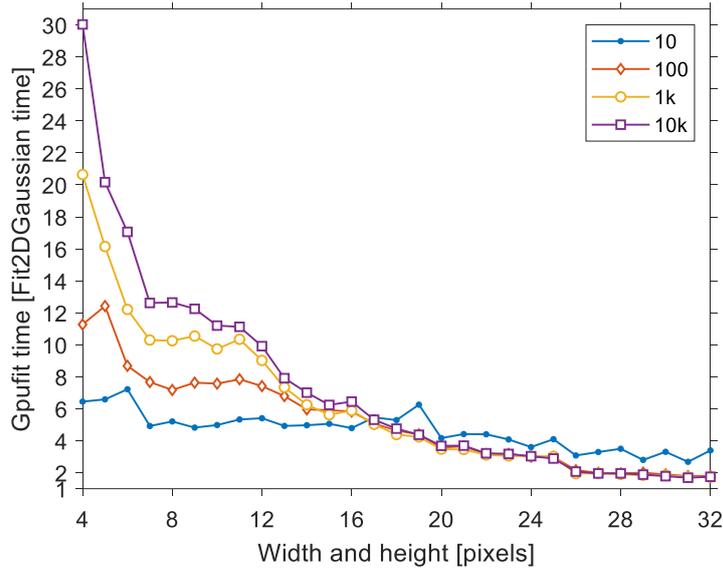

Figure 6. Performance improvement of Fit2DGaussian over Gpufit for batches of 10 to 10'000 fits per call on a NVIDIA Quadro K600 graphics card.

## Accuracy of estimated shaping parameters

To assess the estimation errors, we performed 100'000 fits for typical cases in localization microscopy: image regions of $9 \times 9$ pixels; total signals of $N_\text{signal} = 400$ or 1600 counts with a total background of $N_\text{background} = 40$ counts; and 1600 signal counts without background. Figure 7 compares the absolute errors of the fitted shaping parameters for 400 signal and 40 background counts. The absolute errors are scaled in units of the standard deviation $\sigma_\text{true} \in [1,2]$ pixels of the Gaussian profile. An average error of $\sigma_\text{true}/\sqrt{N_\text{signal}}$ is expected at best without background.

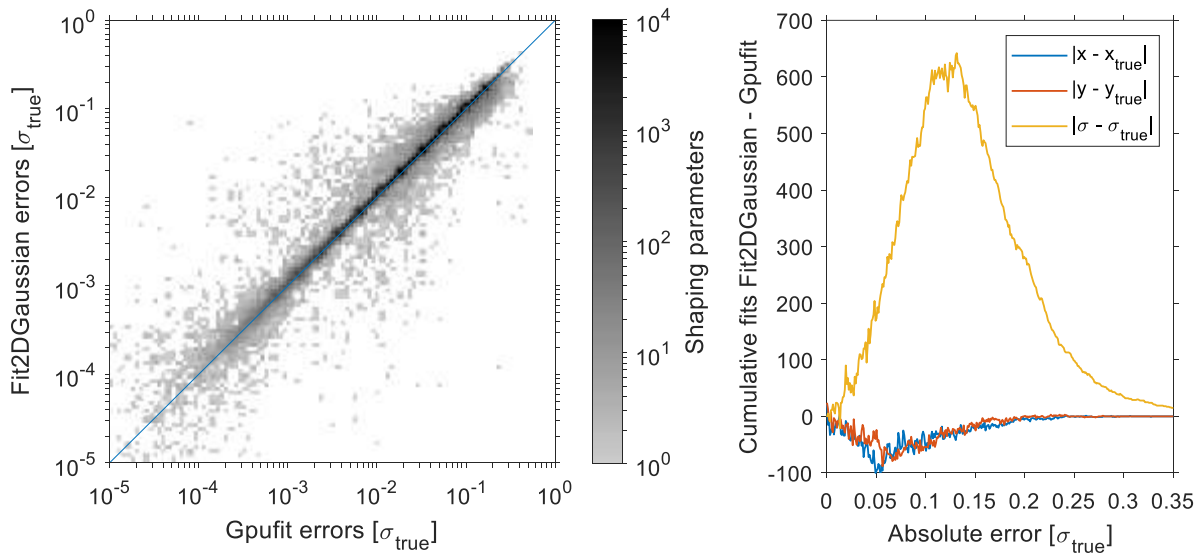

Figure 7. Absolute errors of the shaping parameters $\hat{x}$, $\hat{y}$ and $\hat{\sigma}$ in units of the true PSF size for 100'000 fits. **Left**) Distributions of absolute errors of Fit2DGaussian and Gpufit. **Right**) Difference of the cumulative distributions of the number of fits versus absolute error. Gpufit retrieved $\hat{x}$ and $\hat{y}$ better than Fit2DGaussian in 100 fits. For $\hat{\sigma}$ however, Fit2DGaussian beat Gpufit in 642 fits.

Fit2DGaussian and Gpufit estimated the center coordinates $(\hat{x}, \hat{y})$ accurately and missed the expected average error by 8 to 10 % only. Fit2DGaussian estimated $\hat{\sigma}$ consistently within the expected error as well. Gpufit estimated $|\hat{\sigma}|$ with similar error but negated it for a few percent of the fits and the larger number of fit parameters apparently renders Gpufit more susceptible to find a local error minimum. Table 1 summarizes the results of the error assessments.

Table 1. Median, mean and standard deviation (std) of the absolute fit errors of Fit2DGaussian and Gpufit versus the total signal and background counts.

| $N_{\text{signal}}$ : $N_{\text{background}}$ | Absolute error [$\sigma_{\text{true}}$] | $|\hat{x} - x_{\text{true}}|, |\hat{y} - y_{\text{true}}|$ | | | $||\hat{\sigma}| - \sigma_{\text{true}}|$ | | |
|---|---|---|---|---|---|---|---|
| | | median | mean | std | median | mean | std |
| 400 : 40 counts | **Fit2DGaussian** | 0.0464 | 0.0550 | 0.0418 | 0.0420 | 0.0506 | 0.0396 |
| | **Gpufit(Gauss2D)** | 0.0463 | 0.0550 | 0.0417 | 0.0421 | 0.0514 | 0.0415 |
| 1600 : 40 counts | **Fit2DGaussian** | 0.0228 | 0.0270 | 0.0205 | 0.0203 | 0.0244 | 0.0190 |
| | **Gpufit(Gauss2D)** | 0.0227 | 0.0269 | 0.0205 | 0.0205 | 0.0249 | 0.0201 |
| 1600 : 0 counts | **Fit2DGaussian** | 0.0228 | 0.0269 | 0.0203 | 0.0198 | 0.0238 | 0.0186 |
| | **Gpufit(Gauss2D)** | 0.0227 | 0.0268 | 0.0203 | 0.0203 | 0.0247 | 0.0200 |

## Number of Levenberg-Marquardt iterations

Fit2DGaussian required fewer iterations of the Levenberg-Marquardt algorithm than Gpufit. The saving of Fit2DGaussian is most significant for data with little noise. Figure 8 illustrates the number of iterations performed by both algorithms upon fitting 100'000 images with 1600 signal and 40 background counts. Fit2DGaussian typically terminated after 4 or 5 iterations. Gpufit terminated most often after 5 or 6 iterations but took 8 or 9 iterations frequently as well.

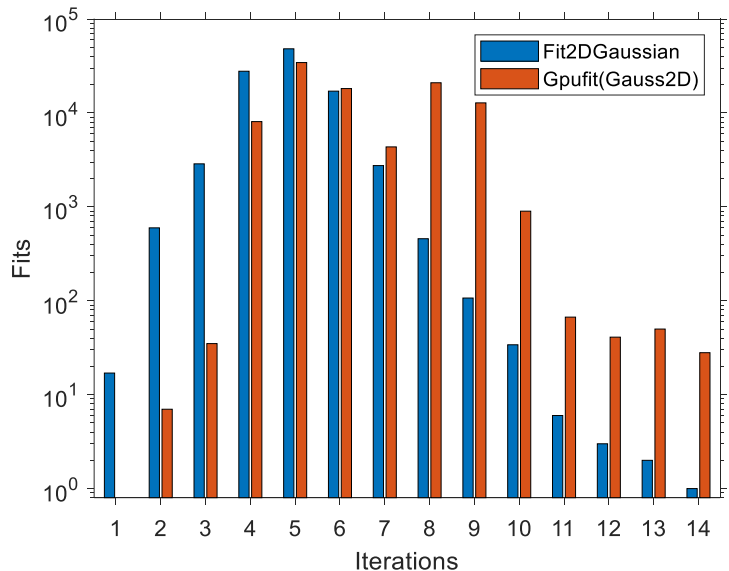

Figure 8. Histogram of the number of executed iterations for 100'000 fits. Fit2DGaussian typically fitted the parameters in five iterations, whereas Gpufit executed five to six or eight to nine iterations.

Fit2DGaussian terminated early mostly because the squared error improved by less than $10^{-6}$. About 15 % of fits terminated because the error did not decrease at all in the last iteration, whereas less than 1 % of fits reached the minimal parameter change criterion. Only very few fits exhausted the number of iterations, sometimes none as in Figure 8. In general, the distribution of stop conditions reflects the signal-to-noise ratio of the images and the criteria values.

# Conclusions

The separation of the amplitude coefficients from the shape-defining parameters allows reducing the number of independent parameters by two when fitting image spots to 2D Gaussian profiles. Explicitly fitting only the shaping parameters increased the sensitivity of the squared error with respect to parameter changes. Hence, the higher calculation effort for the evaluation of the model function paid off by fewer iterations. Even though we did not weight the residuals by the reciprocal of the variance of each pixel value, the modified least squares fit proved robust as it enforces the optimal amplitude coefficients for each shape.

Specializing the algorithm for image regions of interest with at most 1024 pixels allows keeping most intermediate results in GPU registers to reduce slow memory accesses. On a GPU with fast memory, fits of image sizes typically used for the localization of fiducial markers or fluorophores see a two- to three-fold improvement in throughput over Gpufit. On a GPU with slow memory, the improvement was more pronounced. Fit2DGaussian further achieved shorter latency, which supports high control rates of closed-loop sample stabilization.

# Acknowledgments

The authors thank Jan Keller-Findeisen, Adrian Przybylski and Mark Bates for details on the Gpufit implementation and Marco Roose for technical support. We acknowledge funding by the German federal ministry of education and research (BMBF) in the project "New fluorescence labels for protected- and multi-color-STED microscopy (STEDlabel)" (FKZ 13N14122).

## Supporting information

The source code of Fit2DGaussian and the benchmark tools are provided as supplementary material under the General Public License (GPL) version 3.0.

## Supplementary figures

We compared the accuracy of the shaping parameters estimated by Fit2DGaussian and by Gpufit for images sizes of $7 \times 7$, $9 \times 9$ and $11 \times 11$ pixels with signals of $N_{\text{signal}} = 160$, 400 and 1000 counts and 1 background count per pixel. These results are shown in the figures below.

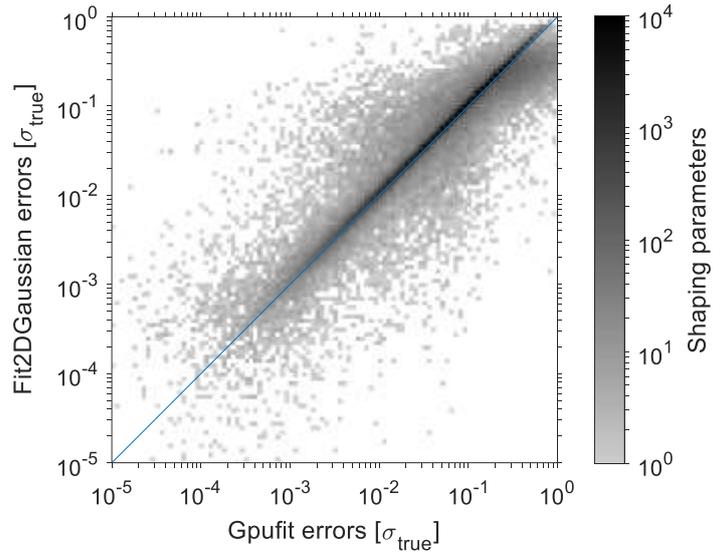

**160 signal and 49 background counts (7×7 pixels)**

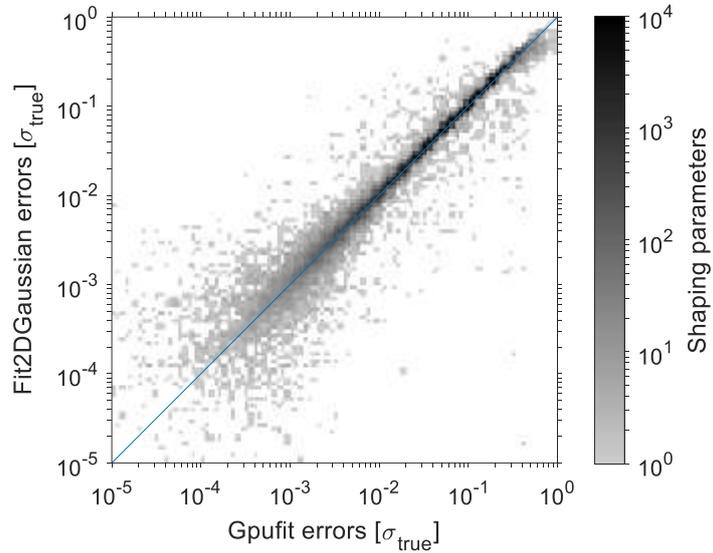

**160 signal and 81 background counts (9×9 pixels)**

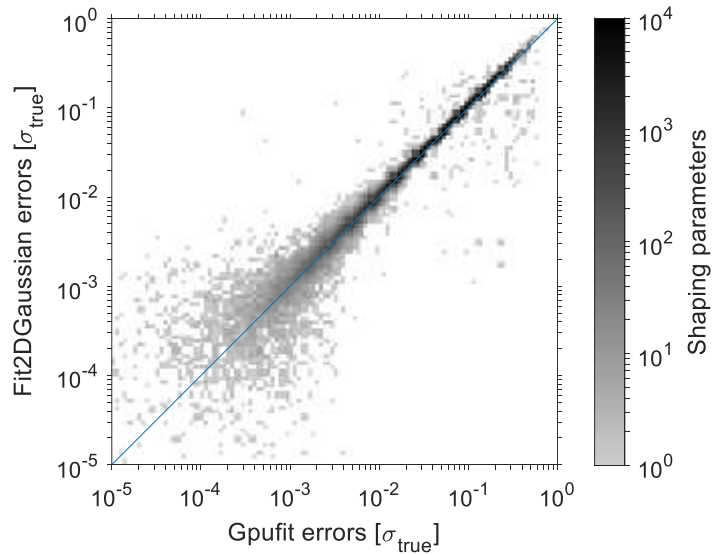

**160 signal and 121 background counts (11×11 pixels)**

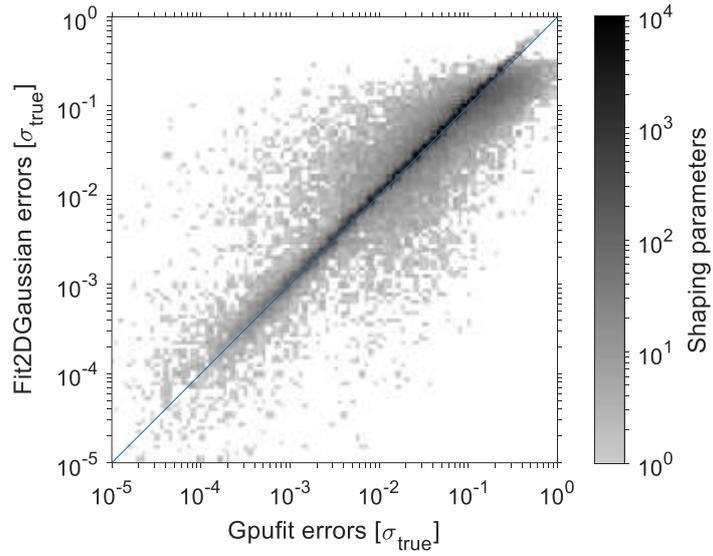

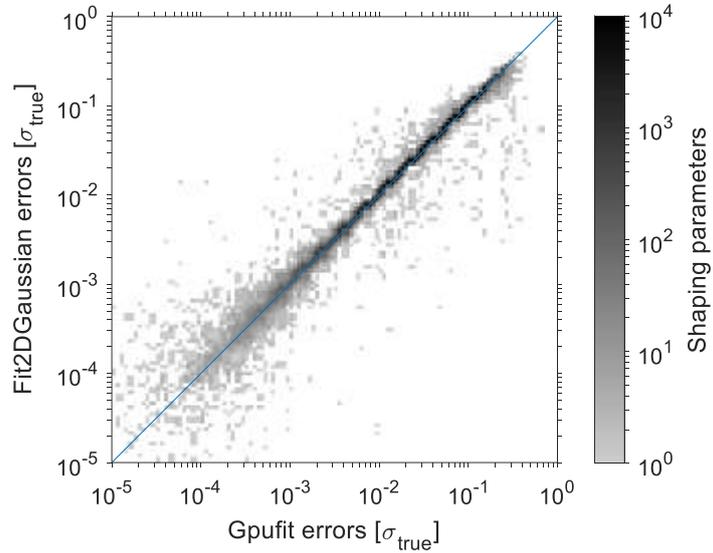

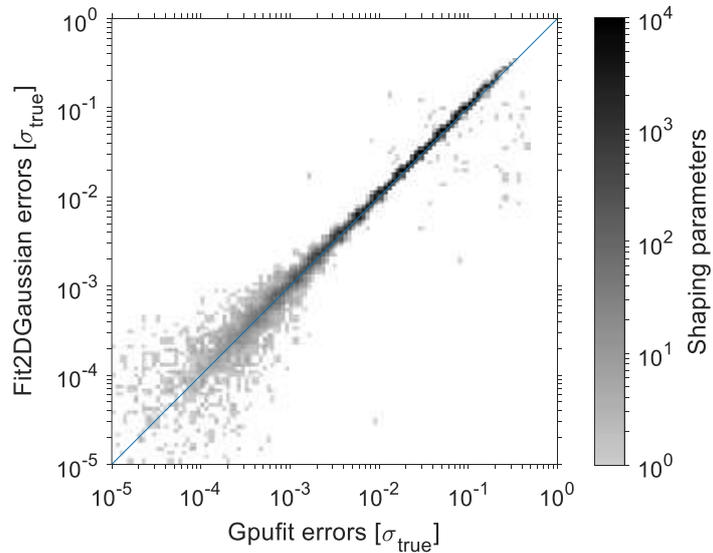

Copyright © Leutenegger & Weber, 3 June 2021                                                                                         Page 15

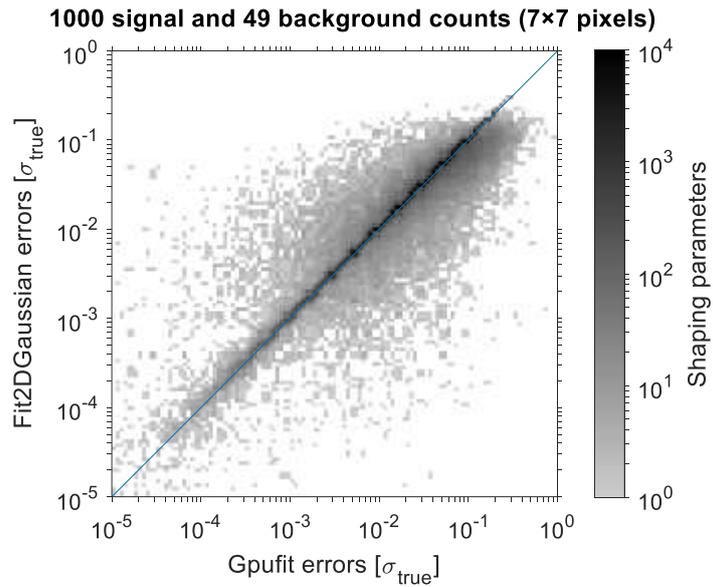
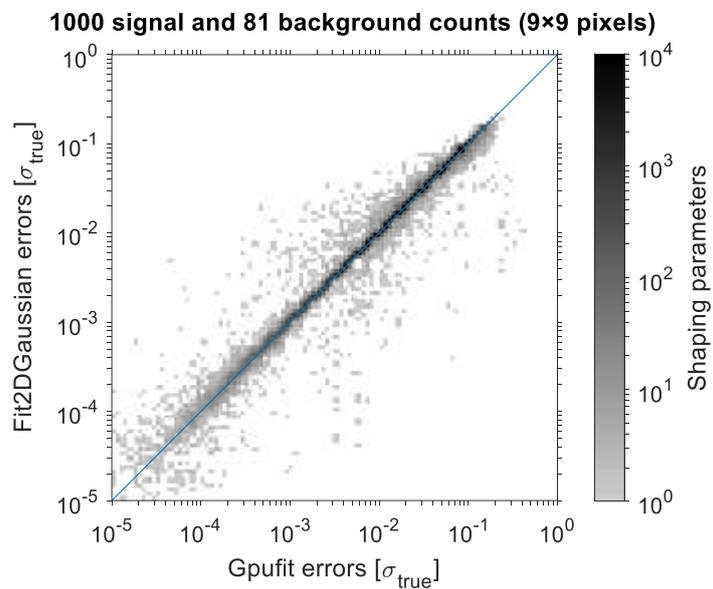
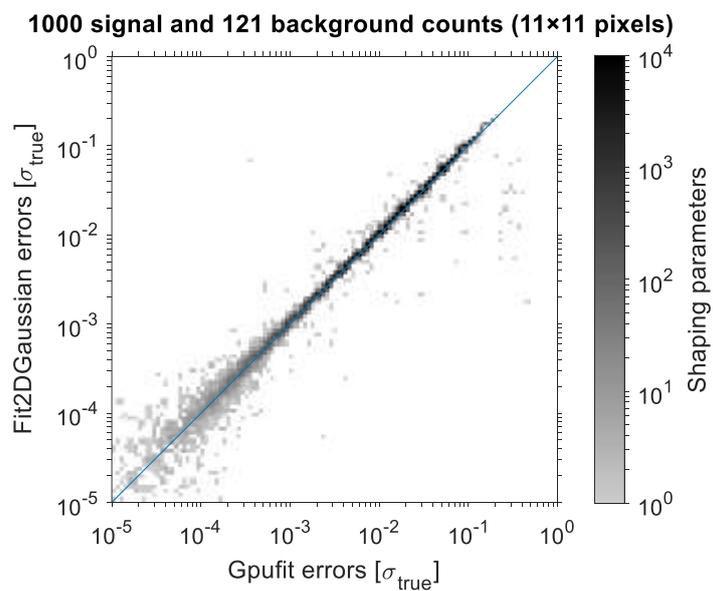